# ALAMOUTI OFDM/OQAM SYSTEMS WITH TIME REVERSAL TECHNIQUE

Ilhem Blel[1] and Ridha Bouallegue[1]

[1]Innov'Com Laboratory Higher School of Communications, University of Carthage, Tunis, Tunisia
`ilhem.bilel@supcom.tn`
`ridha.bouallegue@gmail.com`

## ABSTRACT

*Orthogonal Frequency Division Multiplexing with Offset Quadrature Amplitude Modulation (OFDM/OQAM) is a multicarrier modulation scheme that can be considered as an alternative to the conventional Orthogonal Frequency Division Multiplexing (OFDM) with Cyclic Prefix (CP) for transmission over multipath fading channels. In this paper, we investigate the combination of the OFDM/OQAM with Alamouti system with Time Reversal (TR) technique.*

*TR can be viewed as a precoding scheme which can be combined with OFDM/OQAM and easily carried out in a Multiple Input Single Output (MISO) context such as Alamouti system.*

*We present the simulation results of the performance of OFDM/OQAM system in SISO case compared with the conventional CP-OFDM system and the performance of the combination Alamouti OFDM/OQAM with TR compared to Alamouti CP-OFDM. The performance is derived by computing the Bit Error Rate (BER) as a function of the transmit signal-to-noise ratio (SNR).*

## 1. INTRODUCTION

The use of radio communication systems with multiple transmit and receive antennas also referred to as MIMO system can be used to increase capacity. Because of the time-dispersion that occurs in radio mobile communications, the MIMO channel is frequency selective.

OFDM presents the property to convert such a frequency selective MIMO channel into a set of parallel frequency flat MIMO channels. This makes CP-OFDM a suitable scheme to be associated with MIMO.

Standards such as IEEE802.11a have already implemented the CP-OFDM. Other standards like IEEE802.11n combine CP-OFDM and MIMO in order to increase the bit rate and to provide a better use of the channel spatial diversity.

Nevertheless, CP-OFDM technique causes a loss of spectral efficiency due to the CP as it contains redundant information. Moreover, the rectangular prototype filter used in CP-OFDM has poor frequency localization which makes it difficult for CP-OFDM systems to respect stringent specifications of spectrum masks.

To overcome these drawbacks, OFDM/OQAM was proposed as an alternative approach to CP-OFDM. Indeed, OFDM/OQAM does not need any CP, and it furthermore offers the possibility to use different time-frequency well localized prototype filters such as Isotropic Orthogonal Transform Algorithm (IOTA). One of the characteristics of OFDM/OQAM is that the demodulated transmitted symbols are accompanied by interference terms caused by the neighboring transmitted data in time-frequency domain. The presence of this interference is an issue for some MIMO schemes and until today their combination with OFDM/OQAM remains an open problem.

Some interesting researches [1] [2] [3] propose a modification in the conventional OFDM/OQAM modulation by transmitting complex QAM symbols instead of OQAM ones. This proposal allows to reduce considerably the inherent interference but at the expense of the orthogonality condition. Indeed, the data symbol and the inherent interference term are both complex.

Regarding Alamouti coding, some works have been carried out such as [4] where the authors showed that Alamouti coding can be performed when it is combined with code division multiple access (CDMA). A pseudo-Alamouti scheme was introduced in [5] but at the expense of the spectral efficiency since it requires the appending of a CP to the OFDM/OQAM signal. Another solution was proposed in [6] where the Alamouti coding is performed in a block-wise manner inserting gaps (zero-symbols and pilots) in order to isolate the blocks. The main objective of this paper is to analyze and study the combination of OFDM/OQAM technique with Alamouti system using Time Reversal approach.

Firstly experimented in acoustics and ultrasound domains [7] [8] [9], Time Reversal (TR) has also received attention recently for wireless communications [10] [11] [12]. Owing to its inherent time and spatial focusing properties, TR is now studied as a solution for future green wireless communications [12]. Its time focusing property allows having low intersymbol interferences (ISI) at the receiver side.

In fact, TR was highlighted to be suitable for MISO systems as it is simple prefiltering techniques for any number of transmit antennas and leads to low complexity receivers [13]. Moreover, it reduces the delay spread of the channel [14]. However, to achieve good performance in terms of delay spread reduction and spatial focusing TR must be realized either over a large frequency bandwidth or with multiple antennas [15].

Multicarrier systems such as OFDM are commonly used to deal with time-dispersive channels and can be combined with TR to accommodate any residual intersymbol interference.

The combination of TR and OFDM has recently been studied in [11] [12] and has been proven to allow designing of simple and efficient MISO-OFDM systems [16] [17]. In this paper we investigate the combination of Alamouti OFDM/OQAM with TR.

The remaining of this paper is organized as follow: In section II we describe the OFDM/OQAM modulation in SISO case, and we show that the introduction of appropriate pulse shaping can efficiently combat time and frequency distortions caused by the channel. In section III we present the combination of OFDM/OQAM with multiple transmit antennas and especially with classical Alamouti scheme using TR technique. In section IV we provide simulation results, and finally we give the general conclusions in section V.

## 2. SINGLE-INPUT SINGLE-OUTPUT OFDM-OQAM

The OFDM/OQAM signal in baseband and discrete time for M subcarriers can be expressed, at time kTe, as follows:

$$S_{OFDM/OQAM}[k] = \sum_{m=0}^{M-1} \sum_{n=-\infty}^{+\infty} a_{m,n} \underbrace{f[k - n\frac{M}{2}] e^{j\frac{2\pi}{M}m(k-\frac{LF-1}{2})} e^{j\phi_{m,n}}}_{f_{m,n}[k]} \qquad (1)$$

Where Te denotes the sampling period, $a_{m,n}$ are the real coefficients, f[] is a prototype filter of length LF and $\phi_{m,n}$ denotes a phase term selected for example equal to $\frac{\pi}{2}(m + n)$. Thus, the OFDM/OQAM modulation overcomes the presence of a guard interval or cyclic prefix thanks to a judicious choice of the prototype filter modulating each subcarrier signal which enables well localization in time and frequency, and which verifies the constraint of real orthogonality between subcarriers resulting in:

$\Re\{\langle f_{m,n}, f_{m',n'}\rangle\} = \Re\{\sum_{k=-\infty}^{+\infty} f_{m,n}[k]f^*_{m',n'}[k]\} = \delta_{m,n}\delta_{m',n'}$. Where $\langle g, h \rangle$ denotes the scalar product between g and h. The scalar product $\langle f_{m,n}, f_{m',n'}\rangle$ is a pure imaginary number for $(m, n) \neq (m', n')$. In the following description we use for simplicity the following notation: $\langle f \rangle_{p,q}^{m,n} = -j\langle f_{m,n} f_{p,q}\rangle$.

The prototype filter has to satisfy the orthogonality conditions or at least must be nearly orthogonal. It can be derived directly in continuous-time, as it is the case for instance in [18] with the IOTA filter.

Naturally, the resulting prototype filter has to be truncated and discretized to be implemented. The IOTA prototype filter used in this paper is of length L = 4M and it is denoted by IOTA4. Prototype filters can also be directly derived in discrete time with a fixed length [19]. This is the case of the time frequency localization (TFL) [19] prototype filter. In this paper, it is taken of length L = M and denoted by TFL1.

The block diagram in Figure 1 illustrates our OFDM/OQAM transmission scheme in SISO case. The pre-modulation steps corresponds to a single multiplication by an exponential which argument depends on the phase term Φm,n and on the prototype length. The polyphase block contains the coefficients of the prototype filter. At the receiver side the dual operation are carried out that, at the end taking the real part of the post-demodulator output, allows us to exactly recover the transmitted symbols in the case of a distortion-free channel.

However, in the general case, the propagation channel breaks the real orthogonality condition thus equalization must be performed at the receiver side to restore this orthogonality. Let us consider a channel $h(t, \tau)$ that can also be represented by a complex-valued number $H_{m,n}^{(c)}$ for subcarrier m at symbol time n. At the receiver side, the received signal is the summation of the $S_{OFDM/OQAM}$ signal convolved with the channel impulse response and of a noise component. For a locally invariant channel, we can define a neighborhood, denoted $\Omega_{\Delta m, \Delta n}$ around the (m, n) position, with $\Omega_{\Delta m, \Delta n} = \{(p, q), |p| \leq \Delta m, |q| \leq \Delta n \left| H_{m+p,n+q}^{(c)} \right| \approx H_{m,n}^{(c)}\}$

And we also define $\Omega^*_{\Delta m, \Delta n} = \Omega_{\Delta m, \Delta n} - \{(0,0)\}$

Indeed, despite the use of a prototype filter well localized in time and frequency, OFDM/OQAM modulation produced by building an imaginary term of intrinsic interference. For a SISO system, after transmission over a frequency selective channel and additive noise terms noted η, the demodulated signal can be written as:

$$y_{m,n} = H_{m,n}^{(c)}\left(a_{m,n} + ja_{m,n}^{(i)}\right) + J_{m,n} + \eta_{m,n} \tag{2}$$

$H_{m,n}^{(c)}$ denotes the value of the complex channel on subcarrier m at time n, $\eta_{m,n}$ denotes the noise component at time n and subcarrier m, $ja_{m,n}^{(i)}$ is a purely imaginary term of intrinsic interference affecting the symbol $a_{m,n}$ and dependent on its neighbors symbols at time n given by:

$$a_{m,n}^{(i)} = \sum_{(p,q)\in\Omega_{\Delta m,\Delta n}-(0,0)} a_{m+p,n+q}\langle f \rangle_{p,q}^{m,n} \tag{3}$$

And with $\langle f \rangle_{p,q}^{m,n} = -j\langle f_{m,n} f_{p,q}\rangle$.

$J_{m,n}$, the interference term created by the data symbols outside $\Omega_{\Delta m, \Delta n}$, given by [20]:

$$J_{m,n} = j\sum_{(p,q)\notin\Omega_{\Delta m,\Delta n}} a_{m+p,n+q} H_{m+p,n+q}^{(c)} \langle f \rangle_{p,q}^{m,n} \tag{4}$$

It can be shown that, even for small size neighborhoods, if the prototype functions is well localized in time and frequency, $J_{m,n}$ becomes negligible when compared to the noise term $\eta_{m,n}$. Thus the received signal can be approximated by:

$$y_{m,n} \approx H_{m,n}^{(c)} \left( a_{m,n} + j a_{m,n}^{(i)} \right) + \eta_{m,n} \tag{5}$$

Based on this approximation, various techniques can then be considered at the receiver side to remove the intrinsic interference term $j a_{m,n}^{(i)}$ in case of a SISO system.

If the channel is known at the receiver side, the estimate of the real transmit symbol can be easily obtained by a simple ZF or MMSE [21] equalization as:

$$\hat{a}_{m,n} \approx \Re \left\{ \frac{y_{m,n}}{H_{m,n}^{(c)}} \right\} \approx a_{m,n} + \Re \left\{ \frac{\eta_{m,n}}{H_{m,n}^{(c)}} \right\} \tag{6}$$

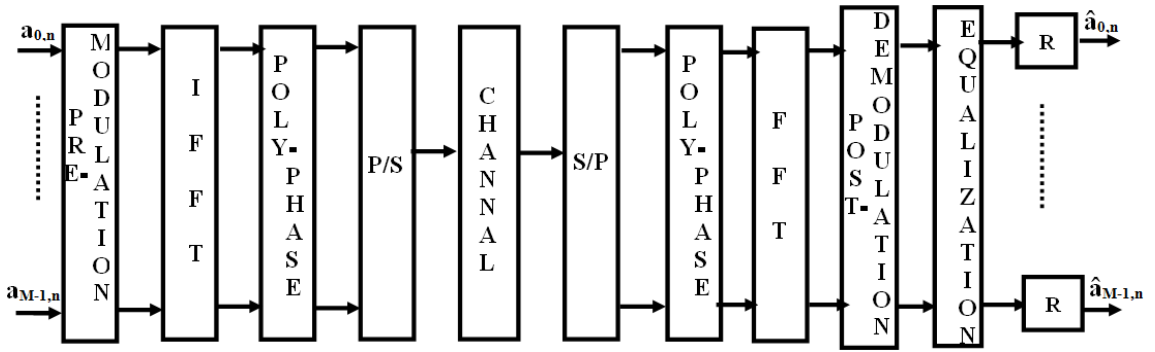

Figure 1. OFDM/OQAM transmission scheme in SISO case

## 3. ALAMOUTI-OFDM/OQAM WITH TIME REVERSAL

Some works [1][2][3] show that when combining OFDM/OQAM with MIMO techniques such as STBC, the presence of the interference term causes problems and makes the detection process very hard if not impossible.

In this section, we shall propose a new Alamouti-OFDM/OQAM scheme in order to get rid of the inherent interference term. Indeed, we will apply the time reversal technique on the outside of the OFDM/OQAM modulator on each transmission antenna.

TR principles are presented in detail for acoustic and electromagnetic waves in [12] and [13] respectively. Applied to wireless communications, TR consists in prefiltering the signal with the time reversed and conjugated version of the channel impulse response (CIR) between transmit and receive antennas. Without loss of generality, such an operation can be represented in discrete time domain as depicted in Figure 2. In this figure, c[l] is the transmit shape filter, h[l] the discrete complex-baseband CIR, h*[−l] the time reversal filter, and c[l] the so-called equivalent channel obtained by convolving h[l] and h*[−l].

Consequently, in the time domain, c[l] is made of a central peak of high amplitude and some side lobes. For rich propagation environments, a time focusing effect is obtained as the channel autocorrelation peak is getting sharper and narrower and as side lobes are reduced. In other words, TR leads to time dispersion compression, here by reducing the ISI occurring between symbols [14].

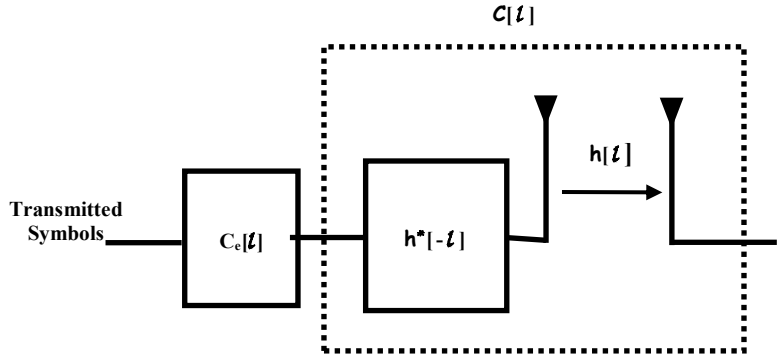

Figure 2. Transmission with Time Reversal technique

If time focusing is sufficiently strong for a given symbol duration, the receiver could merely be reduced to a threshold detector. Nevertheless, for systems exploiting a restricted bandwidth or a limited number of transmit antennas, residual ISI can be efficiently treated through multicarrier approach. The TR and multicarrier approaches can be viewed as complementary and compatible processes applied on the signal before transmission, the former trying to compress the channel time dispersions and the latter accommodating with the residual ISI.

In some preliminary work [16] [17], it has been demonstrated that TR can be applied in an OFDM system either in the time or in the frequency domain. For both implementations, the achieved performance is equivalent.

On that basis, applying TR to an OFDM/OQAM signal amounts to precoding the symbols on each subcarrier by the conjugated channel coefficients. These channel coefficients are obtained from the frequency version of the CIR through a Fourier transform.

More precisely, we proposes to transmit OFDM/OQAM symbols associated with a transmit antenna on an equivalent channel resulting from the convolution of the TR prefilter and the CIR.

Let us suppose that the transmission channel is a frequency selective channel consisting of L paths. We denote $h = (h_0, h_1, ..., h_{L-1})$ the vector of the complex channel coefficients. In the following description, are used indifferently channel coefficients, the temporal impulse response or its transformation in Z to describe and qualify the transmission channel. These various representations are equivalent. It will be the same for the TR prefilter.

These coefficients are distributed according to a centered Gaussian distribution. The transform in Z of the impulse response h(t) of the transmission channel is H(z).

The time reversed from the transmission channel has as a transform $\widehat{H}^*(z^{-1})$ and its impulse response is h*(-t).

The equivalent channel seen by OFDM/OQAM modulation in frequency domain is thus:

$$C(\exp(jw)) = \underbrace{\sum_{l=0}^{L-1} \|h_l\|^2}_{c_0} + 2\sum_{k \neq l} \Re(h_k h_l^*) \cos((k-l)w) - \Im(h_k^* h_l) \sin((k-l)w) \quad (7)$$

The equivalent channel is thus a symmetric conjugate channel which central path c0 is a real coefficient. It also appears, in light of the above equations, that because of the symmetry of the coefficients of the equivalent channel, its transform C(z) is a real function. It will be noted that this reasoning is valid for a discrete or continuous representation of the considered channel. This means that the constraint of real orthogonality between the subcarriers verified by the prototype filter is advantageously verified after passing through this equivalent channel since its equivalent frequency response filter is real, even if the transmission channel associated with each antenna is complex. This avoids the generation of an intrinsic interference term purely imaginary difficult to remove such as that generated for a modulation OFDM/OQAM state of the art.

The diagram in Figure 3 illustrates the principal steps of an OSTBC-MISO-OFDM/OQAM transmission scheme with TR technique. In this paper, we consider the Alamouti system with two transmit antennas and one receiver antenna.

Indeed a sequence of bits is transformed into a sequence of real symbols of QAM constellation. Real symbols are then distributed over M subcarriers and two antennas in a step of serial-parallel conversion, so as to form OFDM/OQAM symbols in the frequency domain for each transmit antenna. There is a real symbol $a_{nm}$ associated with the subcarrier m at time n. The real symbols $a_{nm}$ are then coded at OSTBC coding step according its spatial and temporal dimension to generate two sequences of coded symbols for two transmit antennas. Such codes are described in particular in [20] and in [21] for any number of transmit antennas. In this paper, we consider a space-time coding of real symbols $a_{mn}$, distributed over M subcarriers, defined by the following real orthogonal coding matrix GR2:

$$GR2 = \begin{bmatrix} a_{m,n} & -a_{m,n+1} \\ a_{m,n+1} & a_{m,n} \end{bmatrix}$$

Thus, during the OSTBC coding step, the matrix GR2 is applied to each subcarrier m. Thus obtaining two sequences of coded symbols which are then respectively modulated for the two transmit antennas in accordance with the OFDM/OQAM modulation scheme, and then converted according to an operation of parallel-serial conversion carried out independently for each antenna.

Then, for each Txj antenna, j=1,2, the OFDM/OQAM symbols forming coded OFDM/OQAM multicarrier signal S(j) are filtered by a TRj prefilter defined from an estimate transmission channel.

The vector of the coefficients of the TRj prefilter for the transmit antenna Txj are given by: $h^{TR(j)} = (\hat{h}_{L-1}^{(j)*}, \hat{h}_{L-2}^{(j)*}, \ldots, \hat{h}_0^{(j)*})$ where $\hat{h}^{(j)} = (\hat{h}_0^{(j)}, \hat{h}_1^{(j)}, \ldots, \hat{h}_{L-1}^{(j)})$ is the estimated coefficients of the propagation channel between the transmit antenna Txj and the reception antenna Rx. The resulting multicarrier signal for each antenna is noted:

$$x^{(j)} = s^{(j)} \otimes h^{TR(j)} \tag{8}$$

Multicarrier filtered signals x(j) are then transmitted over the propagation channel via their respective transmit antennas.

Figure 4 illustrates the principal steps of an OSTBC-Alamouti-OFDM/OQAM reception scheme where the received signal equal to the sum of the received signals from each transmit antenna is:

$$y' = \sum_{j=1}^{N} x^{(j)} \otimes h^{(j)} + \eta \tag{9}$$

where η denotes an additive white Gaussian noise.

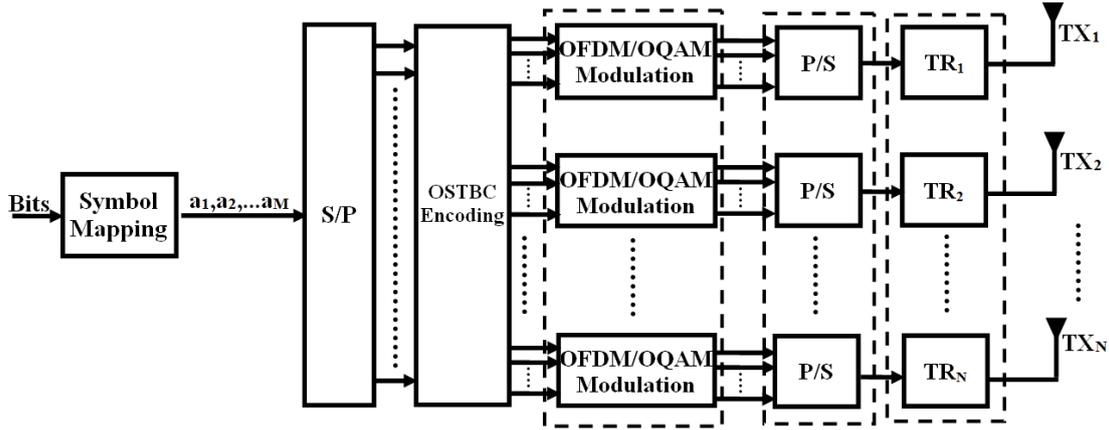

Figure 3. OSTBC-MISO-OFDM/OQAM transmission

The multicarrier signal y', as depicted in Figure 4, are distributed in a serial-parallel conversion of M subcarriers, and then an FFT operation is applied to demodulate OFDM/OQAM symbols.

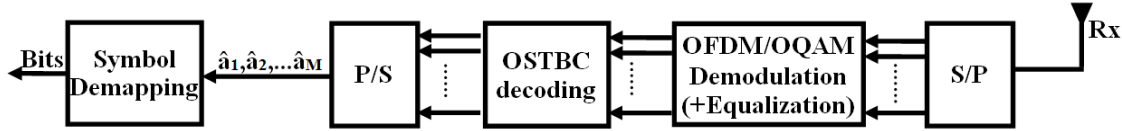

Figure 4. OSTBC-Alamouti-OFDM/OQAM reception scheme

We note HR the equivalent channel obtained by convolution of $h^{TR}$ and h.

With space-time coding GR2 matrix considered previously, the symbols obtained after FFT transformation for each subcarrier m at nT and (n+1)T times are given by the following equations:

$$\begin{aligned} y'_{m,n} &= H^{R(0)}_{m,n} a_{m,n} + H^{R(1)}_{m,n} a_{m,n+1} + \eta_{m,n} \\ y'_{m,n+1} &= -H^{R(0)}_{m,n} a_{m,n+1} + H^{R(1)}_{m,n} a_{m,n} + \eta_{m,n+1} \end{aligned} \qquad (10)$$

Where $\eta_{m,n}$ and $\eta_{m,n+1}$ denote the components of the additive white Gaussian noise for the carrier m at time nT. This expression can also be written in matrix form according to the following expression:

$$\begin{bmatrix} y'_{m,n} \\ y'_{m,n+1} \end{bmatrix} = \underbrace{\begin{bmatrix} H^{R(0)}_{m,n} & H^{R(1)}_{m,n} \\ H^{R(1)}_{m,n} & -H^{R(0)}_{m,n} \end{bmatrix}}_{HC} \begin{bmatrix} a_{m,n} \\ a_{m,n+1} \end{bmatrix} + \begin{bmatrix} \eta_{m,n} \\ \eta_{m,n+1} \end{bmatrix} \qquad (11)$$

Where HC is an orthogonal matrix. An estimate real symbols $\tilde{a}_{mn}$ is thus obtained from the symbols $y'_{m,n}$ resulting from the FFT using the OSTBC decoding for each subcarrier m. After parallel-serial conversion of the real symbols estimated for each subcarrier, the symbols are converted into bits, in accordance with the selected transmission constellation.

Thus, the principle of TR can advantageously eliminate the intrinsic interferences terms generated by the use of the OFDM/OQAM modulation, including MISO system.

## 4. SIMULATIONS RESULTS

In this section, we provide the simulation results of the proposed SISO-OFDM/OQAM system and Alamouti-OSTBC-OFDM/OQAM with TR previously presented with two transmit antenna and one receive antenna.

Our simulations have been carried out with sampling frequency fs=10 MHz; FFT size M = 1024; QPSK modulation is used; Prototype filter IOTA4 and TFL1 is used; 3paths with Power profile (in dB):-0,-3,-2.2 and Delay profile (μ s): 0, 0.2527, 0.32 and finally the ZF equalization technique is used.

The simulations are carried out with a discrete-time signal model and prototype filter of finite length, denoted by L. The first prototype filter is a truncated version of IOTA leading a prototype filter containing L = 4M = 4096 taps, designated by IOTA4. We also use another prototype filter that results from a direct optimization, with L = M = 1024 coefficients, of the time-frequency localization (TFL) criterion [17], designated by TFL1.

As usual, the performance is evaluated by a comparison of the Bit Error Rate (BER) as a function of the SNR ratio.

Figure 5 presents the BER performance comparison between OFDM/OQAM and conventional CP-OFDM systems over multipath channels. Results show that OFDM/OQAM outperforms CP-OFDM. This gain corresponds to the no use of CP.

Figure 6 shows the BER performance comparison between OSTBC-Alamouti-OFDM/OQAM with TR and OSTBC-CP-OFDM system over multipath channels. Simulation results confirm that OSTBC-Alamouti-OFDM/OQAM based TR with IOTA4 prototype filter outperforms Alamouti CP-OFDM. But Alamouti CP-OFDM is better than OSTBC-Alamouti-OFDM/OQAM based TR with TFL1 prototype filter.

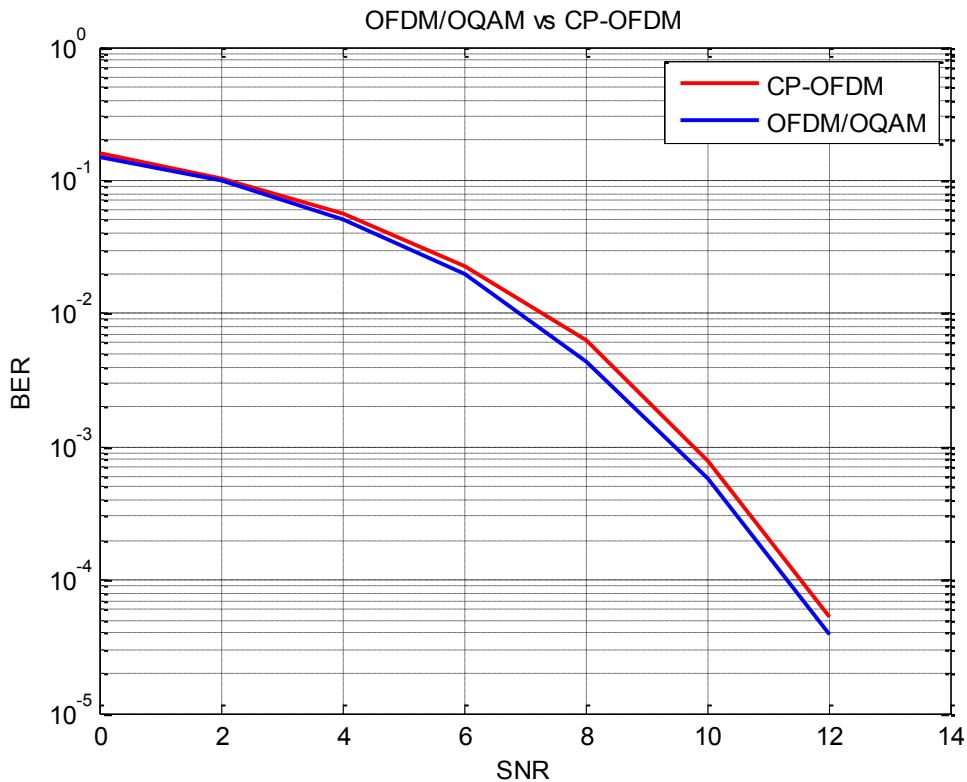

Figure 5. BER performance comparison between OFDM/OQAM and CP-OFDM sytems over multipath fading channel in SISO system

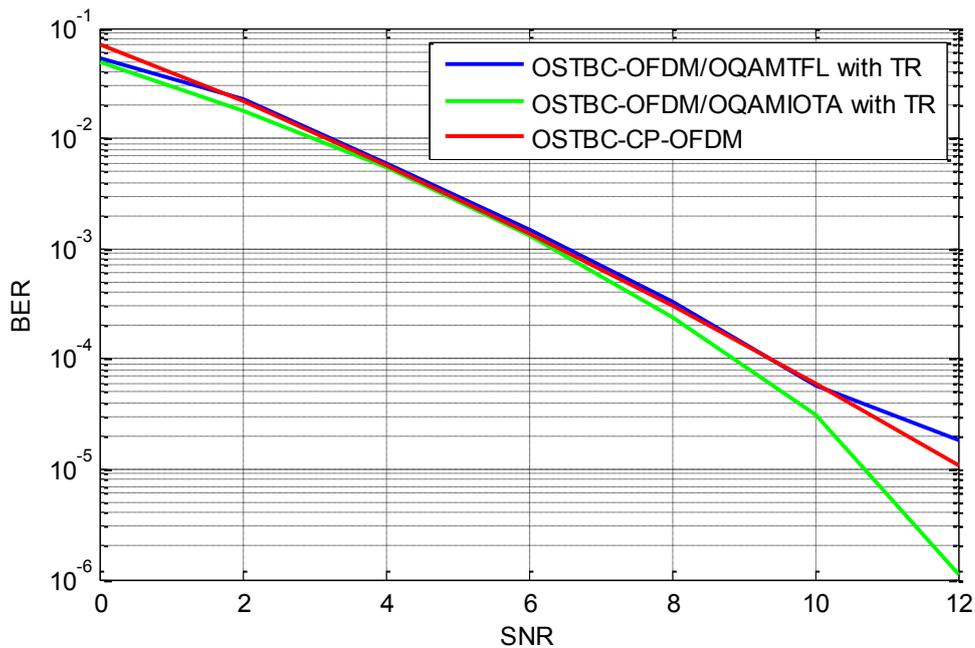

Figure 6. BER performance comparison between OSTBC-OFDM/OQAM with TR and OSTBC-CP-OFDM system with 2×1 Alamouti system

## 5. CONCLUSIONS

This paper addresses the issue of the association of OFDM/OQAM modulation to OSTBC-Alamouti system using TR technique. OFDM/OQAM in a SISO system is presented. OSTBC-Alamouti-OFDM/OQAM-TR is shown to be very simple as the channel becomes purely real in the frequency domain. The receiver only requires a threshold detector in the case of QPSK symbols. This only comes with a TR prefiltering at the transmitter side.

Simulations results prove that OFDM/OQAM performs better than conventional CP-OFDM thanks to the use of special prototype filter such as IOTA and TFL filter and the no use of cyclic prefix. Moreover, OSTBC-Alamouti-OFDM/OQAM based TR with IOTA prototype filter outperforms the OSTBC Alamouti CP-OFDM.

## ACKNOWLEDGMENT

This work was supported by Innov'Com Laboratory located at Higher School of Communications Technological Park of Communications, 2083, Tunis, Tunisia.

**Authors**

**Ilhem BLEL** received the Engineering Degrees from the High School of Communications of Tunis (SUP'COM) in 2007. In Juin 2009, she received the master degree on electronic and telecommunications systems from National Engineer School of Monastir ( ENIM). Since January 2009, she has worked as a university assistant in the high Institute of Mathematic and Informatic of Monastir (ISIMM). She is currently working toward the Ph.D. degree in Telecommunications systems at the High School of Communications of Tunis in Innov'Com laboraotory.

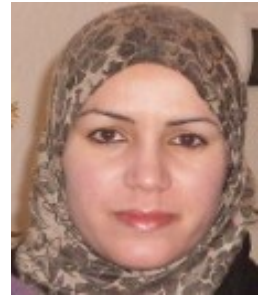

**Ridha BOUALLEGUE** received the Ph.D degrees in electronic engineering from the National Engineering School of Tunis. In Mars 2003, he received the Hd.R degrees in multiuser detection in wireless communications. From September 1990 he was a graduate Professor in the higher school of communications of Tunis (SUP'COM), he has taught courses in communications and electronics. From 2005 to 2008, he was the Director of the National engineering school of Sousse. In 2006, he was a member of the national committee of science technology. Since 2005, he has been the Innov'COM laboratory research in telecommunication Director's at SUP'COM. From 2005, he served as a member of the scientific committee of validation of thesis and Hd.R in the higher engineering school of Tunis. His current research interests include wireless and mobile communications, OFDM, space-time processing for wireless systems, multiuser detection, wireless multimedia communications, and CDMA systems.

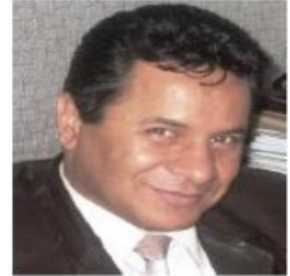